\documentclass[aps,pra]{revtex4}
\usepackage{amsmath}
\usepackage{amsfonts}
\usepackage{amssymb}
\usepackage{graphicx}

\begin{document}

\title{Two-qubit parametric amplifier: large amplification of weak signals}

\author{S. Savel'ev$^{1,2}$, A.M. Zagoskin$^{1,2}$, A.L. Rakhmanov$^{3,2,1}$, A.N. Omelyanchouk$^{2,4}$, 
Z. Washington$^1$, Franco Nori$^{2,5}$}
\affiliation{$^1$Department of Physics, Loughborough University, Leicestershire, LE11 3TU, United Kingdom\\$^2$Advanced Science Institute,
RIKEN, Wako-shi, Saitama 351-0198, Japan\\
$^3$Institute for Theoretical and Applied Electrodynamics, Russian Academy of Sciences, 125412 Moscow, Russia\\
$^4$B.Verkin Institute for Low Temperature Physics and Engineering,  61103, Kharkov, Ukraine\\
$^5$Physics Department,
The University of Michigan, Ann Arbor, MI 48109-1040, USA}

\begin{abstract}
 Using numerical simulations, we show that two coupled qubits can
 amplify a weak signal about hundredfold. This can be achieved
 if the two qubits are biased simultaneously by this weak signal
 and a strong pump signal, both of which having frequencies close
 to the inter-level transitions in the system.
The weak signal strongly affects the spectrum generated by the
strong pumping drive by producing and controlling mixed harmonics
with amplitudes of the order of the main harmonic of the strong
drive.
We show that the amplification is robust with respect to noise, with an intensity
of the order of the weak signal. When deviating from the optimal regime (corresponding to
strong qubit coupling and a weak-signal frequency equal to the inter-level transition
frequency) the proposed amplifier becomes less efficient, but 
it can still considerably enhance a weak signal (by several tens).  
 We therefore propose to use coupled qubits as a combined
parametric amplifier and frequency shifter.
\end{abstract}
\maketitle

\section{introduction}
While the desire of building code-cracking quantum computers \cite{Bellac} remains as elusive as ever since its inception, its pursuit has been one of the major contributing factors to the enormous progress achieved in quantum mesoscopic physics and quantum nanodevices over the last decade and a half. These efforts have already resulted in the development of a new branch of mesoscopic digital and analogue devices \cite{Nori-new,Nori1,Nori2,buluta,smirnov,mi} and even new types of materials known as quantum 
metamaterials \cite{metamat,ost-zag}, allowing to control quantum coherent media. In this article we describe how two coupled
qubits can be used as a parametric amplifier. 

\subsection{Superconducting qubits}


%
From the point of view of quantum computing, a qubit is any
two-state quantum system which satisfies certain control and
readout requirements and can therefore be used for the execution
of quantum algorithms (see, e.g., \cite{Bellac}, Ch.~2). Many
efforts have been devoted to the study of theoretical quantum
information processing and, more recently, significant progress
has been achieved on experimental aspects of this field. For
example, one of the most fascinating results obtained in
experimental mesoscopic physics over the past decade has been the
realization of several types of superconducting qubits, which
demonstrated many of the qualities required for quantum
information processing. Moreover, superconducting qubits in
quantum electronics have a potentially much wider range of
applications than just quantum information processing (see, e.g.,
%
%
Refs.~[\onlinecite{Nori-new,Nori1,metamat,Nori2,Girvin2009,Wendin2006,Zagoskin2007}]),
including: single-photon generators, producing quantum (squeezed
or Fock) states of the electromagnetic field, quantum transmission
lines, quantum amplifiers, etc. Various types of superconducting
qubits have been produced, including charge, flux, and phase
qubits. These use different physical mechanisms to control their
states and store information. For example, in a charge qubit, the
state $|1\rangle$ has one extra Cooper pair, compared to the state
$|0\rangle$, while for a flux qubit, the two logical states differ
by a fraction of the magnetic flux quantum. Superconducting qubits
are mesoscopic (e.g., their working quantum states may differ by
dozens of millions of single-particle states) and scalable (i.e.,
it is possible to link these together). Moreover, their
fabrication, control, and readout require techniques already well
developed in solid state electronics.

Several superconducting qubit designs of various degrees of complexity and performance have been realized
(reviewed in, e.g., Refs.~\onlinecite{Nori-new,Nori1,buluta,Girvin2009,Wendin2006,Zagoskin2007}). 
The so-called persistent-current flux qubit \cite{Mooij,Orlando1999} combines relative simplicity with decent decoherence times and scalability. It consists of a small superconducting loop (approximately 10 $\mu$m across) interrupted by three Josephson junctions. The flux quantization condition is
\begin{equation}
\phi_1 + \phi_2 + \phi_3 + 2\pi\Phi/\Phi_0 = 2\pi n,
\label{eq:eq1}
\end{equation}
where $\phi_j$ is the phase difference across the $j$th junction, $\Phi$ is the total magnetic flux through the loop, $\Phi_0 = h/2e$, and $n$ is an integer. This allows to eliminate one of the phases (e.g., $\phi_3$). Due to the small self-inductance of the loop, the difference between the flux $\Phi$ and the external flux $\Phi_x$, as well as the magnetic energy of the system,  can be neglected. On the other hand, one must include the contribution to the energy from the charges on the Josephson junctions, $Q_j^2/2C_j$,  and the Josephson energy $-E_j \cos \phi_j$, where $C_j$ and $E_j$ is the capacitance and the Josephson energy of the $j$th junction, respectively. Up to numerical factors, the junction charges are the momenta canonically conjugate to the  Josephson phase differences (see, e.g., \cite{Zagoskin2011}, \S 2.3). Introducing variables $\phi_{\pm} = (\phi_1 \pm \phi_2)/2$, it is straightforward to show that the classical Hamilton function of the system (assuming that two junctions are identical, $C_1 = C_2 = C, E_1 = E_2 = E$, and that $C_3 = \alpha C, E_3 = \alpha E$) is given by
\begin{equation}
{\cal H} = \frac{\Pi_+^2}{2M_+} + \frac{\Pi_-^2}{2M_-} - E \left[ 2 \cos \phi_+ \cos \phi_- + \alpha \cos \left(2\phi_-+2\pi\frac{\Phi_x}{\Phi_0}\right)\right],
\label{eq:eq-ham}
\end{equation}
where $\Pi_{\pm}$ are the corresponding momenta and $M_{\pm}$ are determined by the junction capacitances. The potential energy term in the interval is periodic and, for the proper choice of $\alpha$ and with $\Phi_x \approx \Phi_0/2$, it contains a double well, the minima of which are almost degenerate and correspond to the current flowing (counter)clockwise around the loop. These states are chosen as physical qubit states.  Transitions between them are enabled by quantum tunneling through the barrier. The quantum-mechanical Hamiltonian of the persistent-current qubit is obtained from (\ref{eq:eq-ham}) by using the relation $\hat{N} = -i\partial/\partial \phi$ between the superconducting phase and the number of extra Cooper pairs (proportional to the charge on the Josephson junction). Its detailed analysis is given in \cite{Mooij,Orlando1999}. Truncating the Hilbert space of the system to  the two lowest states, which can be done due to the strong anharmonicity of the potential in (\ref{eq:eq-ham}), its Hamiltonian can be reduced to the standard pseudospin form,
\begin{equation}
H = -\frac{1}{2} \left( \epsilon\; \sigma_z + \Delta\; \sigma_x \right).
\label{eq:pseudospin}
\end{equation}
Here $\epsilon$ is proportional to the external flux through the qubit, and $\Delta$ is determined by the tunneling matrix element between the potential minima.  Here we note that persistent-current flux qubits have decoherence times in excess of $10\:\mu$s at the
operating frequency $\sim 1$ GHz, allow successful coherent coupling of
several qubits, and show steady improvements. Therefore, these
 superconducting qubits are promising new elements for versatile
quantum circuits.

\subsection{Parametric amplifiers}
A parametric amplifier is based on the idea of a parametric resonance occurring for a linear oscillator 
with parameters oscillating in time. Such an amplifier \cite{param} is implemented as a mixer, where an input weak signal is mixed 
with a strong local oscillator signal producing the strong output.

In recent years, several new mesoscopic systems have been proposed and implemented as parametric amplifiers. These
systems include small molecules in intense laser fields \cite{molec}, polaritons in semiconductor microcavities \cite{cavities},
current-voltage oscillations in SQUIDs \cite{squid} and Josephson junctions \cite{Joseph,robert,paul}. 
Another proposal uses
 active and tuneable metamaterials \cite{param-meta} to amplify weak signals. Most of these proposals wish to develop
a very compact parametric amplifier, which can even demonstrate quantum amplification \cite{quant-ampl}. This indicates that
quantum qubit systems (and, in particular, superconducting qubit systems) could be very promising candidates 
for mesoscopic parametric amplifiers.  


\subsection{Outline of results}

The main results of this work can be summarized as follows.
\begin{itemize}
\item Consider an ac drive $A\sin (\omega t)$ applied to a
four-level quantum system (e.g., a couple of flux qubits (e.g.,
\onlinecite{nocoupl,coupl}) placed in an ac magnetic field, see Fig.~\ref{fig2}) with
$\omega_{\rm pump} = \omega$ in resonance with a transition
between a pair of its energy levels. Even though the evolution of
the density matrix is described by a linear equation,  the
spectrum of the density matrix elements (Fig.~\ref{fig1bis}) shows
a sequence of peaks corresponding to different harmonics of the ac
drive (which is commonly thought to occur only in nonlinear
systems). More interestingly, the strongest oscillations are not
at the main frequency $\omega$, but at some of its harmonics,
forming a hierarchy of resonances, which is unusual for nonlinear
systems. The explanation of this phenomenon lies in the structure
of the master equations' set for the density matrix elements, in
which the external signal enters multiplicatively rather than
additively.
\item Such a {\it hierarchy of parametric resonances\/} makes this
two qubit system (e.g., two coupled flux qubit) 
an efficient parametric amplifier and frequency shifter,
especially if the weak signal has its frequency $\omega_{\rm weak}
= \tilde{\omega}$ close to another inter-level transition of the
system. The latter can be achieved by tuning the qubit coupling
(e.g., like in Ref.~\onlinecite{coupl}, see also Fig.~\ref{fig2}
below). In this case, a weak signal $\epsilon\cos \tilde{\omega}t$
generates a combination of harmonics $m\omega+k{\tilde \omega}$.
The amplitudes of these harmonics increase with
$\epsilon$, and can reach a height of the order of the main
harmonic of the drive, even for very small $\epsilon$. Thus, the
weak signal can be amplified by a factor of up to several hundred.
\item The amplification effects (Fig.~\ref{fig2bis}) are not suppressed by a realistic
amount of decoherence (dephasing and relaxation) in the system.
Thus, e.g., currently available superconducting flux qubits with
relatively short coherence times can be used as nanoscale,
coherent amplifiers of weak signals in the frequency range of
several hundred MHz.
\item Noise (Fig.~\ref{fig4bis}) of the order of the signal cannot suppress 
signal amplification (Fig.~\ref{fig5bis}), which is also robust (but, of course, weaker) 
with respect to both: changing the frequency of the weak signal (Fig.~\ref{fig6bis}a) and 
the parameters of the coupled-qubit system (Fig.~\ref{fig6bis}a).
\end{itemize}

\section{Quantum amplification with qubits}

One of the challenging tasks for which superconducting qubits seem
to be well suited is the amplification of a weak signal, a
crucially important tool for both technological and scientific applications.
 For this goal, different
types of linear or nonlinear resonance devices are commonly used
(e.g., Ref.~\onlinecite{param,rakh}). The problem of signal
amplification becomes very difficult at the nano-scale. Here we
demonstrate that two coupled qubits can be employed as a
parametric amplifier \cite{param} based on the effect of the
parametric resonance between a weak signal and quantum
oscillations between the quantum levels of the system, driven by
an external ac signal (the pump signal). While the actual
realization details of the qubits is irrelevant for the
mathematics of the problem, we stress that the implementation of
the proposed scheme is ideally suited for mesoscopic
superconducting qubits because these are very controllable and
versatile.

Two coupled qubits can be described by the Hamiltonian:
\begin{equation}
    H = -\frac{1}{2} \sum_{j=1,2} \left[ \Delta_j\; \sigma^j_z + \epsilon_j(t)\;\sigma^j_x \right] + g\; \sigma^1_x\: \sigma^2_x
    \label{eq-ham}
\end{equation}
where $\sigma^j_z$ and $\sigma^j_x$ are Pauli matrices
corresponding to either the first ($j=1$) or the second ($j=2$)
qubit;  the eigenstates of $\sigma^j_z$ are the basis states of
the $j$th qubit at zero coupling.

\begin{figure}[btp]
\begin{center}
\includegraphics*[width=10.0cm]{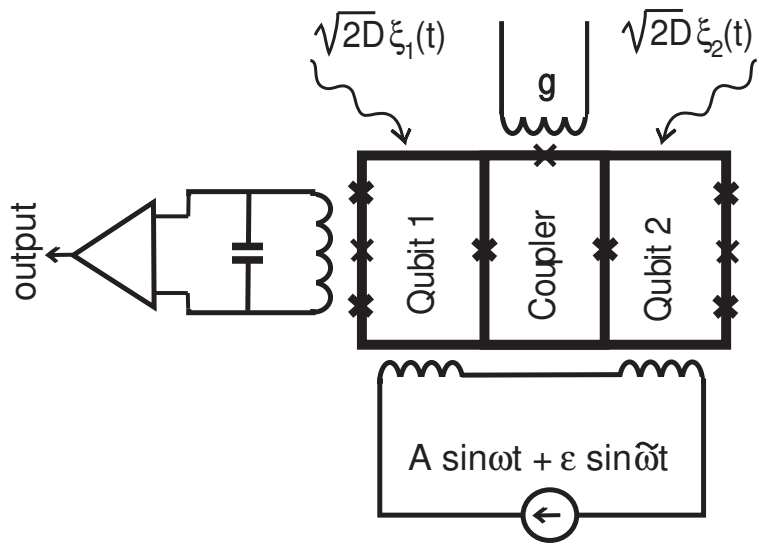}
\end{center}
\caption{ Schematic diagram of two  flux qubits (persistent
current qubits) coupled via a coupler loop. Josephson junctions
are represented by crosses, and their thickness indicate their
Josephson critical currents (drawn not to scale).  The qubit
states and the amplitude and sign of the coupling constant $g$ are
controlled by the corresponding magnetic fluxes, $f_e^{(1,2)}$ and
$f_{\rm coupler}$ (in  units of $\Phi_0$). Both the pumping drive and the weak 
signal $A\sin\omega t+\epsilon \sin {\tilde \omega} t$ are generated by a source 
in the bottom circuit. The left circuit is needed to pick up an output signal $Z_1(t)$. The
top circuit controls the coupling $g$. Noise, shown by $\sqrt{2D}\xi_i(t)$, is coupled to 
each qubit.}\label{fig2}
\end{figure}

To be more specific, let us consider two coupled
nominally-identical superconducting persistent-current flux qubits
(e.g., Ref.~\onlinecite{Mooij}), where each of the latter consist of a superconducting
loop interrupted by three Josephson junctions, along which
persistent-currents can circulate controlled by applied magnetic
fluxes. When two such loops are placed next to each other, so that
they feel each other's magnetic fields, this situation naturally
produces the ``antiferromagnetic'' coupling represented in Eq.~(\ref{eq-ham}) 
by the $\sigma_x$--$\sigma_x$ term, with $g > 0$ (see,
e.g., Ref.~\onlinecite{nocoupl}). More elaborate designs can produce a
tunable coupling (see, e.g.,
Refs.~\onlinecite{coupl,coupling-refs}), with the amplitude and
sign of $g$ controlled externally by the magnetic flux, $\Phi_{\rm
coupl}$, through the coupler loop \cite{coupl} (see
Fig.~\ref{fig2}).

The state of each qubit is controlled by the magnetic flux
$\Phi_e^{(j)} = f_e^{(j)}(t)\, \Phi_0$ through it, where $\Phi_0 =
h/2e$ is the flux quantum. In the vicinity of $f_e^{(1)} =
f_e^{(2)} = 1/2$, the ground state of each qubit is a symmetric
superposition of states $|L\rangle$ and $|R\rangle$ with,
respectively, clock- and counterclockwise circulating
superconducting currents of the same magnitude $I_p$. In the basis
$\{|L\rangle$, $|R\rangle\}^{(1)} \otimes \{|L\rangle$,
$|R\rangle\}^{(2)}$ the two-qubit system can be described by the
four-level Hamiltonian (\ref{eq-ham}) with $\epsilon_j =
I_p\Phi_0\delta\!f_e^{(j)}$; here $\delta\!f_e^{(j)}(t) $ contains
both the pump and the input signals. The tunnelling amplitudes
$\Delta_j$ are usually fixed by the fabrication process, but can
be tuned if one of the qubit junctions is replaced by two
junctions in parallel, in a dc SQUID configuration (see, e.g.,
\cite{you,Paauw}). The interaction constant $g$, as mentioned
above, can be made tuneable using a coupler loop
(Fig.~\ref{fig2}). Note that the density matrix spectrum is
directly related to the immediately measurable current/voltage
spectrum in the tank, which was exploited in \cite{i1} to detect
Rabi oscillations in a flux qubit.

For simplicity, we consider two identical qubits; that is, we
assume $\Delta_1=\Delta_2=\Delta$. A direct diagonalization of the
Hamiltonian (\ref{eq-ham}) leads to  the inter-level transition
frequencies
\begin{eqnarray}
\begin{array}{lll}
  \omega^{(1)} = 2\sqrt{\Delta^2+g^2}, & & \omega^{(2)} = \sqrt{\Delta^2+g^2}- g,
 \\
\omega^{(3)} = \sqrt{\Delta^2+g^2} + g, & &  \omega^{(4)} = 2g,
\end{array} \label{omegaS}
\end{eqnarray}
which are tuneable by changing $g$. Therefore, $\omega^{(1)},\, ...,\, \omega^{(4)}$ can be adjusted to a desirable frequency, that 
was used below.

Let us drive the qubits simultaneously by a control ac pump
signal, with frequency $\omega_{\rm pump} = \omega$
and amplitude $A$, as well as a weak input signal with frequency
$\omega_{\rm weak} = \tilde{\omega}$ and amplitude $\epsilon\ll
A$,
to be amplified:
\begin{equation}
\epsilon_j(t)=A\sin(\omega t)+\epsilon\sin({\tilde{\omega}} t) + \sqrt{2D}\xi_j(t)
\label{drive1}
\end{equation}
where we also take into account a noise term (with intensity $D$) which can be due to fluctuations in the signals or 
an environmental noise. As an example, we consider white noise with zero mean: $\langle\xi(t)\rangle=0$ and
$\langle \xi_j(t)\xi_l(t')=\delta_{jl}\delta(t-t')$, where $\delta$ refers to either the Dirac delta function
or the Kronecker delta. 
The qubit density matrix $\hat{\rho}$ can be written as
\begin{equation}
\hat{\rho}=\frac{1}{4} \sum_{a,b=0,x,y,z} \Pi_{ab} \: \sigma^1_a \otimes \sigma^2_b. \label{master}
\end{equation}
This is a straightforward generalization of the standard representation of the single-qubit density matrix expression using the Bloch vector; the components $\Pi_{ab}$ thus constitute what can be called the Bloch tensor. Then, the master equation,
\begin{equation}
\frac{d\hat{\rho}}{dt} = -i\left[\hat{H}(t),\hat{\rho}\right]+\hat{\Gamma}\hat{\rho},
\label{poxyj}
\end{equation}
can be written down directly [see Eq.~(\ref{poxy}) in the Appendix],
 using the standard approximation for the dissipation operator $\hat{\Gamma}$ via the
 dephasing $(\Gamma_{\phi 1}, \Gamma_{\phi2})$, and relaxation $(\Gamma_{1}, \Gamma_{2})$
 rates, to characterize the intrinsic noise in the system. Also, for simplicity,
 hereafter we assume that the relaxation rates are the same for both identical
 qubits, i.e., $\Gamma_{\phi 1} = \Gamma_{\phi 2}=\Gamma_{\phi}$ and
 $ \Gamma_{1}=\Gamma_{2}=\Gamma_{r}$, and the temperature is low enough,
 resulting in $ Z_{T2}= Z_{T1}=1$, where $Z_{Tj} = \tanh(\Delta_j/2k_BT_j)$
 is the equilibrium value of the $z$-component of the Bloch vector.

In the limit of zero coupling, $g = 0$, there exists a solution of
Eqs.~(\ref{poxy}) with no entanglement between the qubits. This
solution can be written as a direct product of two independent
density matrices expressed through their Bloch vectors:
\begin{equation}
\hat{\rho}_j = \frac{1}{2} (1+X_j\sigma_x+Y_j\sigma_y+Z_j\sigma_z).
\end{equation}
 The components of the Bloch tensor $\Pi_{ab}$ are all zero with the exception of
 \begin{equation}
(\Pi_{ox},\Pi_{oy},\Pi_{oz})= (X_1,Y_1,Z_1); \:\: (\Pi_{xo},\Pi_{yo},\Pi_{zo})=(X_2,Y_2,Z_2)\; ,
\label{eq:qomponents}
\end{equation}
which are just the separate qubits' Bloch vector components. If
the interaction is nonzero, $g \neq 0$, the entanglement between
these qubits makes the components of the Bloch tensor nonzero, and
such an entangled state can persist for some time even if the
interaction is switched off later on.

\section{Simulation results}

\begin{figure}[btp]
\begin{center}
\includegraphics*[width=14.0cm]{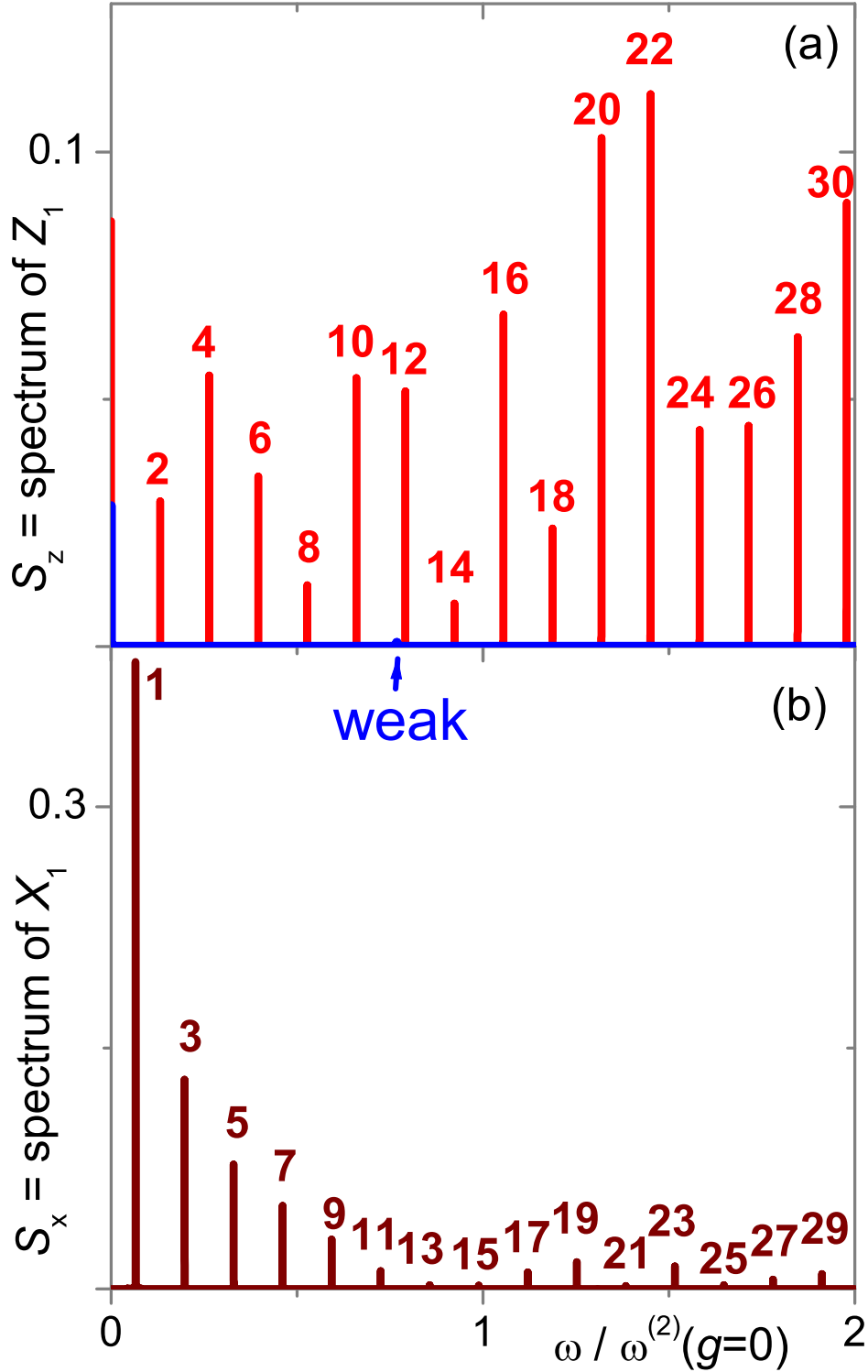}
\end{center}
\caption{(Color online) The spectrum $S_Z$ of the $Z_1$ matrix
element (responsible for the occupation of the excited level of
the first qubit) is shown for dephasing and relaxation given by
$\Gamma_{\phi}/\Delta=\Gamma_{r}/\Delta=10^{-3}$ (we use the same 
$\Gamma_{\phi}$ and $\Gamma_r$ for all results reported in this article). (a) When
the weak signal amplitude is equal to zero ($\epsilon = 0$) and
when the reduced-drive amplitude $A/\Delta=15$, the ac
monochromatic drive fed into the two-qubit amplifier is converted
into a set of even harmonics $\omega_{m}=m\omega$, with $m=2, 4,
6...$. Note the absence of response at $\omega_{1}=\omega$ (no peak
there). For comparison, $S_Z$ is shown in blue and indicated by the
blue arrow for zero-drive amplitude $A=0$ and weak signal
$\epsilon/\Delta=0.1$. Only one very small peak is hardly seen,
indicated by the blue arrow, corresponding to $2\omega_{\rm weak}$.
(b) Spectrum $S_X$ of the off-diagonal matrix element
$X_1$ (zero signal case). The same ac monochromatic drive fed into the amplifier is
converted into a set of odd harmonics for $S_X$: $m = 1, 3, 5,
\dots$.
 Note that (a) and (b) show very unusual non-monotonic spectra. The
 observed non-monotonicity in the system response can be seen as
 a fingerprint of the qubits, characterizing their dynamical
 nonlinear response.}\label{fig1bis}
\end{figure}

\begin{figure}[btp]
\begin{center}
\includegraphics*[width=14.0cm]{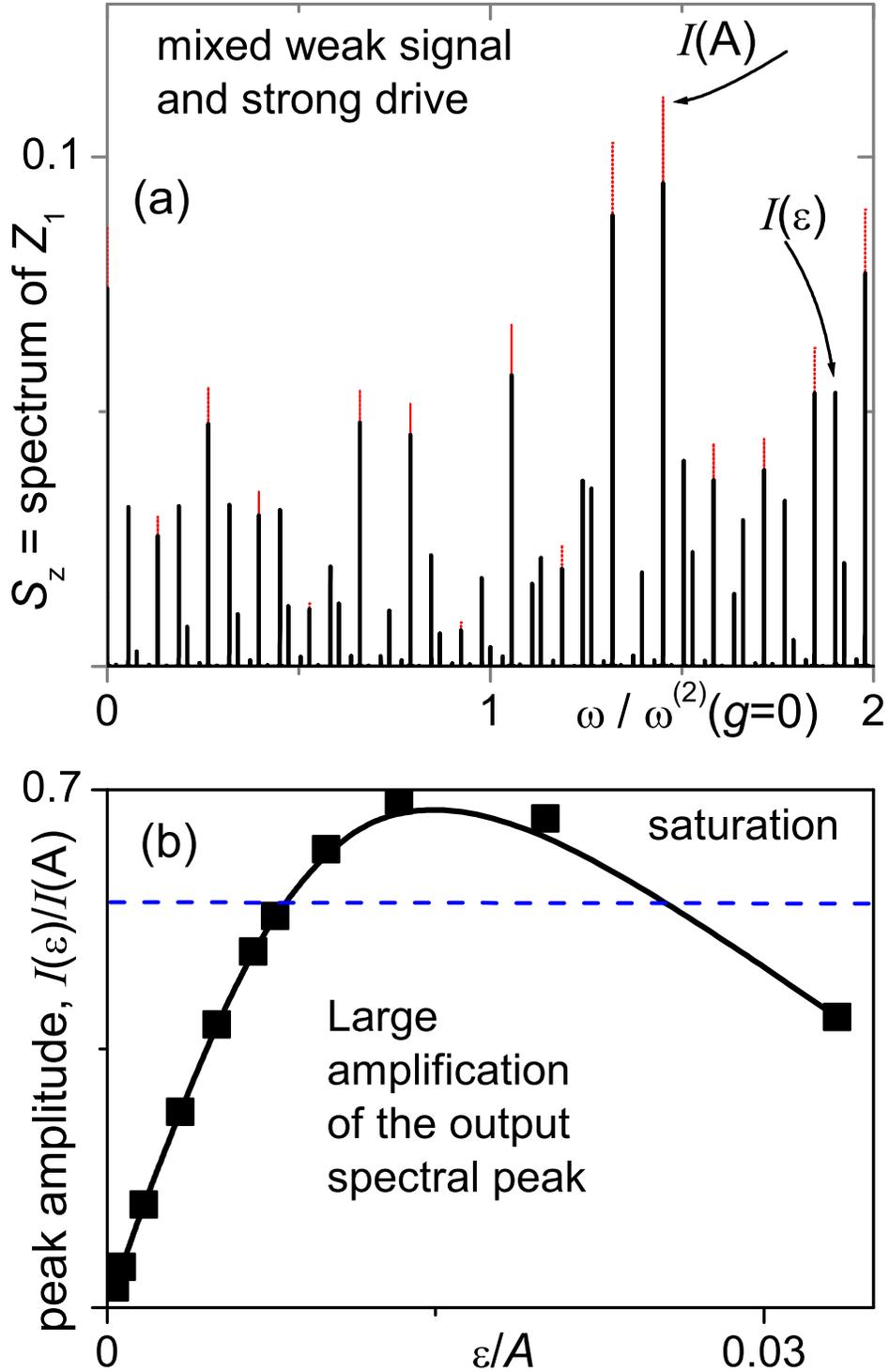}
\end{center}
\caption{(Color online) (a) The amplification of a weak signal
$\epsilon/\Delta=0.1$ by a strong drive $A/\Delta=15$ (i.e., $\epsilon/A=1/150$) can be seen
in the spectrum of $S_Z$ shown by the black solid lines. The spectrum obtained is {\it not a simple
superposition} of the two  spectra $S_Z(A/\Delta=15,\epsilon=0)$ (also shown here by red vertical
dotted peaks) and $S_Z(A=0, \epsilon/\Delta=0.1)$ [see also red and blue peaks in Fig.~\ref{fig1bis}(a)].
Instead, a combination of harmonics appears for
$k\omega+l{\tilde{\omega}}$, with integer $k$ and $l$. The height
of these peaks is almost proportional to $\beta\epsilon$, for
$0<\epsilon/A<0.005$, and it is strongly enhanced by a factor
$\beta\sim 100$. The enhanced peak is marked by an arrow with symbol $I(\epsilon)$. Note
that this peak is absent in Fig.~\ref{fig1bis}(a). 
(b) Normalized output amplitude $I(\epsilon)/I(A)$ (ratio of
the heights of the highest mixed peak $I(\epsilon)$ to the highest peak $I(A)$ of the spectrum at $\epsilon=0$) 
as a
function of the reduced weak signal amplitude
$\epsilon/A$. This dependence shows an almost-linear increase of
the peak height with $\epsilon$ for small $\epsilon/A$, followed by saturation and even decay
at relatively large $\epsilon/A$.}\label{fig2bis}
\end{figure}

\begin{figure}[btp]
\begin{center}
\includegraphics*[width=14.0cm]{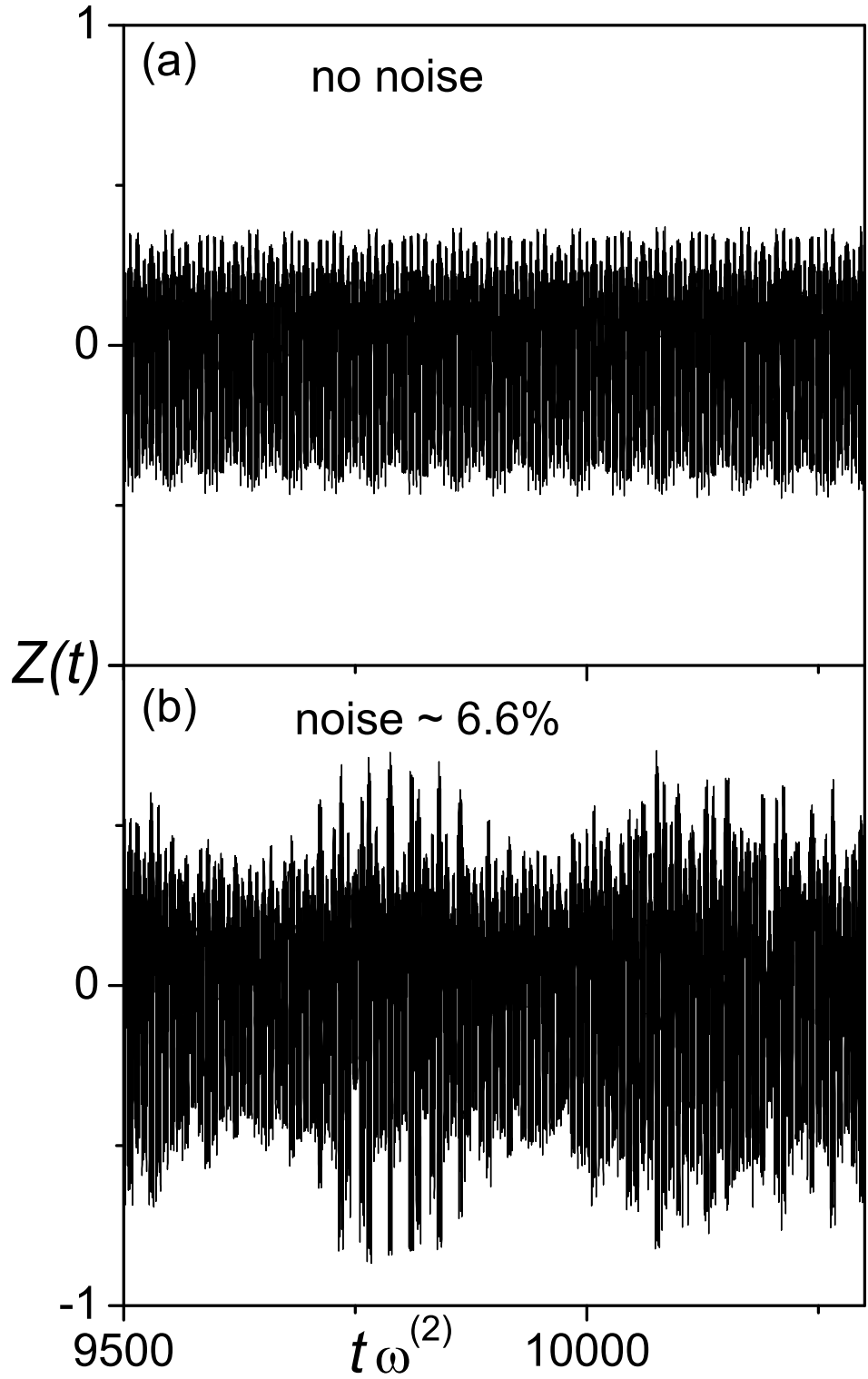}
\end{center}
\caption{(Color online) Trajectories $Z_1(t)$ for different noise levels:
$\sqrt{D}/\epsilon=0$ for (a) and $\sqrt{D}/\epsilon=0.066$ for (b). All other
parameters are the same as in Fig.~\ref{fig2bis}a. As seen from (b), the applied white
noise considerably affects the time dependence of $Z_1(t)$, making trajectories quite
noisy with respect to the noiseless situation shown in (a).}\label{fig4bis}
\end{figure}

\begin{figure}[btp]
\begin{center}
\includegraphics*[width=14.0cm]{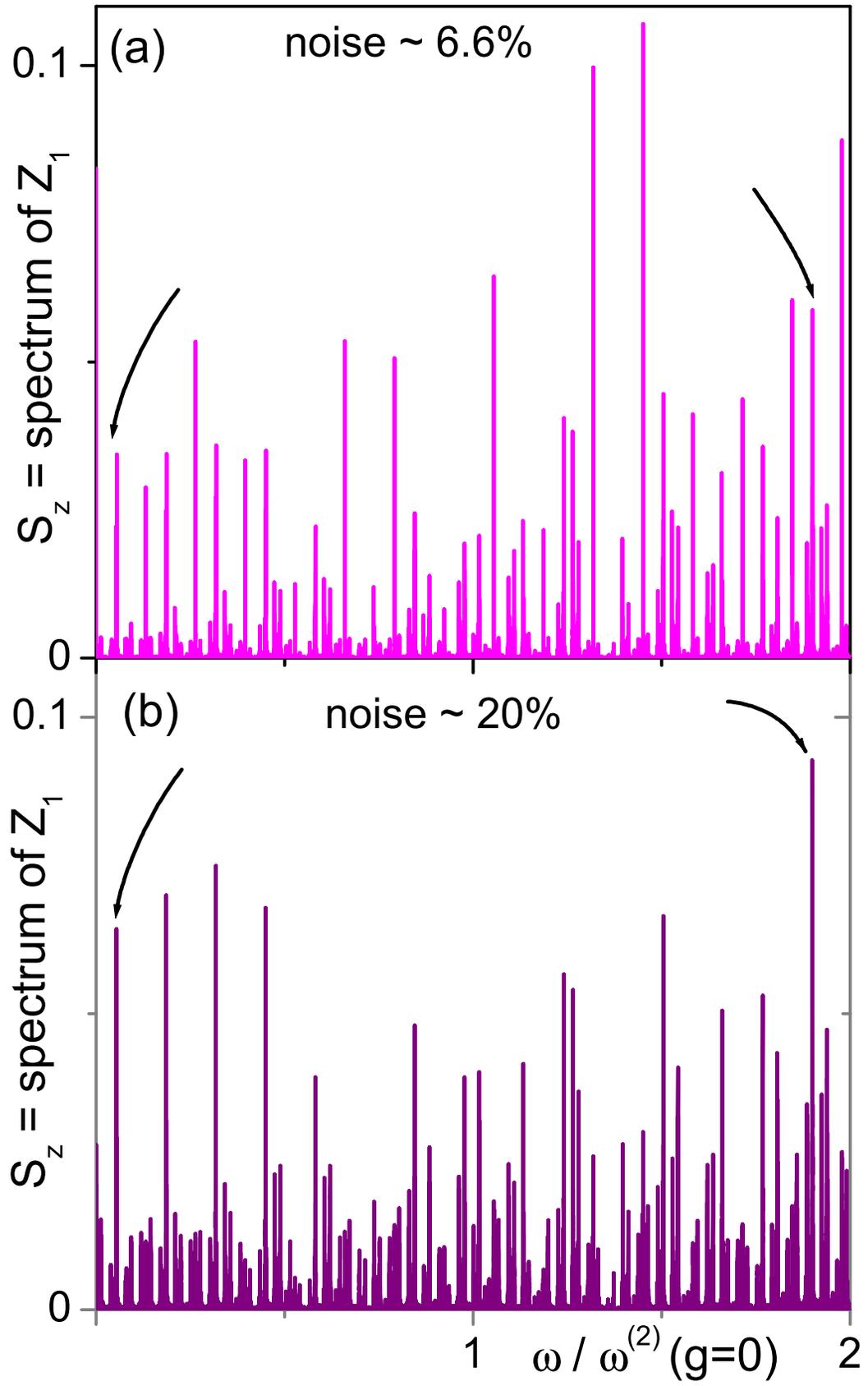}
\end{center}
\caption{(Color online) Spectrum $S_z(\omega)$ for the same parameters used in Fig.~\ref{fig2bis} and
different noise levels $\sqrt{D}/\epsilon=0.066$ for (a) and 0.2 for (b). One can clearly see
the mixed (pump-signal) peaks for several combination-frequencies
$k\omega_{\rm pump}+l\omega_{\rm weak}$, even at high noise levels, when the trajectories
$Z_1(t)$ shown in Fig.~\ref{fig4bis} are very noisy.}\label{fig5bis}
\end{figure}

\begin{figure}[btp]
\begin{center}
\includegraphics*[width=14.0cm]{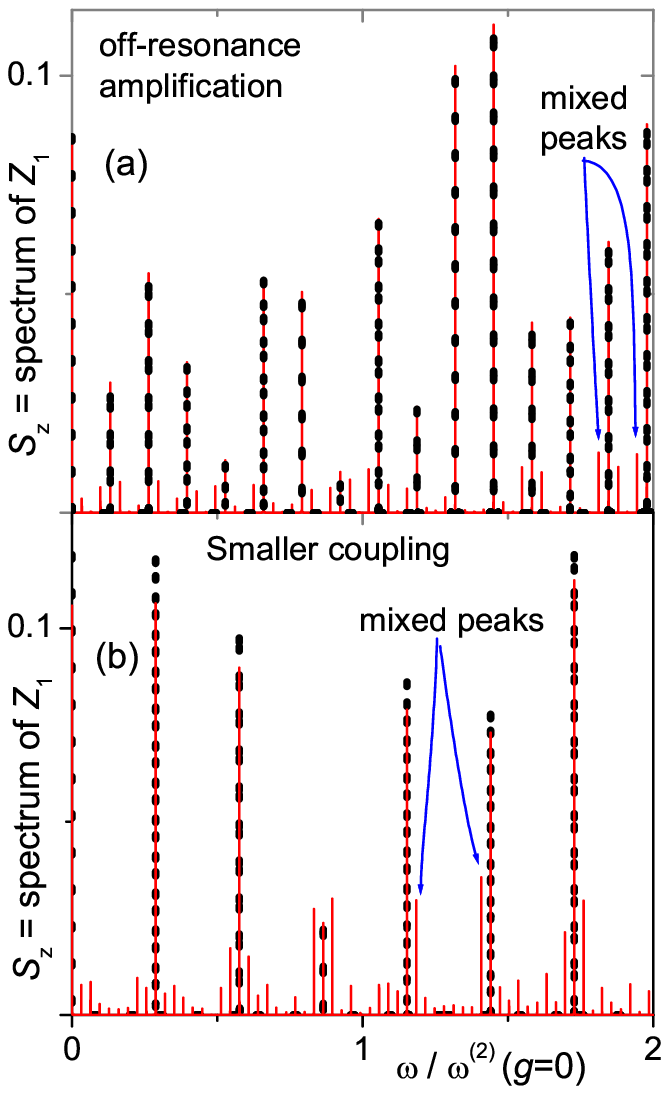}
\end{center}
\caption{(Color online) Amplification of a weak signal away from the optimal regime. 
(a) Spectrum $S_z$ for off-resonance frequency $\omega_{\rm weak}=1.113\omega^{(3)}(g=1)$
of a weak signal shown by red thin solid line. All other parameters of the simulations are
the same as in Fig.~\ref{fig2bis}. The thick black dotted line corresponds to the simulations
with zero signal, thus, highlighting the combination-frequency amplified harmonics as red lines
with no black dots on top. One can see that the amplification of off-resonance
signal is weaker, but still remarkable with amplification factor $\beta$ about 30. (b) Spectrum
of $S_z$ for weakly coupled qubits $g/\Delta=0.1$ and weak signal amplitude $\epsilon/A=24$, as shown by
the red thin solid line; all other parameters are as in Fig.~\ref{fig2bis} and $A/\Delta=12$. The same simulations but with zero signal 
is shown by a thick black dotted line to highlight the amplified mixed peaks. The estimated amplification
factor $\beta$ is about 10.}\label{fig6bis}
\end{figure}

Formally, the set of equations (\ref{poxy}) might look complicated
but these are just a set of fifteen coupled ordinary differential
equations which can be easily integrated numerically
\cite{mdbook}. Indeed, measuring time in units of $1/\Delta$, we
numerically solved Eqs.~(\ref{poxy}) by the Euler method for two
coupled qubits, driven by the field (\ref{drive1}) with
$\omega_{\rm pump} = \omega = \omega^{(2)}$ and $\omega_{\rm weak}
= {\tilde{\omega}} = \omega^{(3)}$ [see Eq.~(\ref{omegaS})]. Our
numerical integration produces a set of data for $\Pi_{a,b}(t)$
which can be analyzed using a Fourier transform. Thus, we can
numerically evaluate the spectrum of $Z_1=\Pi_{oz}$, defined as
\begin{equation}
S_Z(\omega) = \langle |Z_1(\omega)| \rangle,
\label{eq:SPECTRAS}
\end{equation}
and the similarly defined spectrum of $X_1 = \Pi_{0x}$. 
If one considers a flux qubit as a particular realization of our
two-qubit parametric amplifier, then $Z_1(t)$ and $X_1(t)$ (and, thus, their spectra
$S_Z$ and $S_X$) can be directly obtained by using impedance measurements \cite{ilich} of 
the circulating current in the first
qubit of the proposed device.
These
spectra are shown in Fig.~\ref{fig1bis} for $g/\Delta=1$. Below we
will analyze the influence of a weak signal on $S_Z$ and $S_X$.

When the weak signal is switched off [Fig.~\ref{fig1bis}(a)], the
spectrum of $Z_1$ exhibits several peaks corresponding to the
harmonics, $\omega_{m}=m\omega$, of the applied drive ($A/\Delta=15$). In contrast
to the standard nonlinear response, where the spectrum starts from
$\omega_{1}$ and has harmonic peaks $\omega_{m}$, whose heights
{\it decrease} with $m$, in our two-qubit system, $S_z$ starts
with $\omega_{2}$ and contains only even harmonics. The peak heights 
show a surprising {\it
non-monotonic} dependence on $m$, with the two highest peaks
occurring at $m=20$ and $m=22$. A somewhat similar non-monotonic
peak dependence on the harmonic number is seen for $S_X$, the
spectrum of the off-diagonal matrix element $X_1=\Pi_{ox}$. Even
though this spectrum starts with the main harmonic $\omega_1$, it contains
only odd peaks which heights still show a non-monotonic dependence on $m$
[Fig.~\ref{fig1bis}(b)]. The super-indices in $\omega^{(1)},
\omega^{(1)}$, etc., refer to the level-splitting (intrinsic
properties) of the qubits, while the lower indices in $\omega_1,
\omega_2, \dots$ label the harmonics of the response to the input
ac monochromatic drive. We measure all frequencies in units of the energy-splitting
of independent qubits $\omega^{(2)}(g=0)$, since this provides a characteristic frequency 
in the system which is fixed even if one tunes the inter-level frequencies $\omega^{(i)}(g)$.

All spectral peak heights are proportional to the drive amplitude
$A$. Thus, a weak signal should produce a spectrum with strongly-suppressed 
peaks. This is consistent with our simulations for
zero-drive amplitude $A=0$ and  weak-signal amplitude
$\epsilon/\Delta=0.1$. The resulting spectrum has only one peak,
with an amplitude about 150 times lower than the highest peak of
the $S_Z$ spectrum for $A/\Delta=15,\epsilon=0$
[Fig.~\ref{fig1bis}(a), the weak peak is shown there as ``weak'' 
and it is almost invisible].

When mixing the strong drive with a weak signal, the resulting
spectrum $S_Z$ at $A/\Delta=15,\epsilon/\Delta=0.1$ (i.e., $\epsilon/A=1/150$)
[Fig.~\ref{fig2bis}(a)] strongly differs from the simple
superposition of the two spectra described above,
$S_Z(A/\Delta=15,\epsilon=0)$ and $S_Z(A=0,\epsilon/\Delta=0.1)$.
It is remarkable that {\it the weak signal produces a considerable
change in the spectrum:} it generates peaks at combination
frequencies, $k\omega+l{\tilde{\omega}}$, with integer $k$ and
$l$; their heights are determined by the weak signal but can be of
the order of the highest peak in $S_Z(A/\Delta=15,\epsilon=0)$. 
Interestingly, the enhancement of combination-frequency harmonics occurs by 
borrowing some energy from the pumping drive, which own harmonics (harmonics
existing at $\epsilon=0$) decay when the weak signal is applied. Indeed, if one compares 
the spectrum at $\epsilon=0$ shown by red dotted lines, with the spectrum at $\epsilon/\Delta=0.1$ shown
by black solid lines in Fig.~\ref{fig2bis}a, one can see that the heights of the harmonics at
$\epsilon=0$ are higher that the corresponding peaks at $\epsilon/\Delta=0.1$, even though the total
energy pumped in the system is larger when both the pumping drive and the weak signal are applied.

The heights of the combination-frequency peaks increase with the weak signal
amplitude, $\epsilon$, followed by saturation at large values of $\epsilon$.
Of course, the heights of the combination-frequency peaks tend to zero when $\epsilon\rightarrow 0$.
It is clearly seen [Fig.~\ref{fig2bis}(b)] that their height can
be approximated by a linear function, $\beta\epsilon$, for the
weak signal amplitude in the range $0<\epsilon/A<0.005$, with an
{\it amplification coefficient\/} $\beta$ {\it of about\/} 100.
This level of amplification is remarkable.

It is also useful to stress that many peaks associated with different combinations of
two frequencies, $k\omega_{\rm pump}+l\omega_{\rm weak}$, 
appear in the spectrum of $S_Z$ for different integers $k$ and $l$.
In other words, the spectrum has many combination-frequency harmonics with different intensities. 
This allows to pick up the signal
on a frequency which better fits an available experimental setup. Thus, the proposed {\it two-qubit amplifier
is also a frequency shifter}, allowing to shift the frequency of a weak signal to a desirable frequency range. 
Note that it is usually hard to predict which peak should be the highest one [see Fig.~\ref{fig2bis}
with highest peak marked as $I(\epsilon)$].
However, by measuring the output signal $I_{\pm}$ of the first mixing harmonics, $\omega_{\rm weak}\pm\omega_{\rm pump}$,
usually  allows to pick up a strongly amplified signal. For instance, the ratio of the highest peak $I_{\epsilon}$
to $I_{\pm}$ is about 1.7 in our simulations shown in Fig.~\ref{fig2bis}a. Therefore, choosing the harmonics 
$\omega_{\rm weak}\pm\omega_{\rm pump}$ would be a good guide to observe the predicted signal amplification.   

\section{Noise influence}

Since a weak signal can be considerably amplified by a strong drive in our parametric amplifier,
one can wonder if an uncontrollable noise could also be amplified making the weak signal
indistinguishable. To check this we performed simulations at different noise levels chosing all other
parameters as in Fig.~\ref{fig2bis}, i.e., $A/\Delta=15,$ and $\epsilon/A=1/150$. A noise with intensity 
of the order of 6.6\% with respect to the weak signal (i.e., $\sqrt{D}/\epsilon\sim 0.066$) has already considerably 
affected the time dependence of the measured signal (see, Fig.~\ref{fig4bis}), but it does not strongly 
influence the spectrum of $S_Z(\omega)$, as seen in Fig.~\ref{fig5bis}a. 
The reason for this is that the noise contains all frequencies
(or at least a broad frequency spectrum), thus, its energy pumped in the signal harmonic is relatively small.

Even stronger noise, $\sqrt{D}/\epsilon=0.2$, is still not enough to suppress the peaks attributed to the weak signal.
Moreover, a sort of stochastic resonance (increase of the peak heights with noise) \cite{chem-phys} is also seen on Fig.~~\ref{fig5bis}b.
Therefore, the proposed two-qubit parametric amplifier is robust with respect to noise and, moreover, the noise 
can even be used to further amplify the signal.   

\section{Away from the optimal regime: robust amplification of two-qubit amplifier}

In order to realize the strongest amplification of a weak signal (optimal working regime), a tuneable 
coupling must be used. In other words,
the coupling should be tuned to adjust a level splitting frequency $\omega^{(i)}$  
to the frequency $\omega_{\rm weak}$ of the applied weak signal. This
adjustment cannot be always realized. Thus, the amplification of a weak signal of an arbitrary frequency 
(different from the level splitting) is worth studying. As we expected, the amplification of the weak signal on the frequency
 $\omega_{\rm weak}=1.113\omega^{(3)}\ne \omega^{(i)}$ is weaker (Fig.~\ref{fig5bis}a)
 by a factor of about 3.5 (ratio of the highest combination-frequency peaks in Figs.~\ref{fig2bis} and \ref{fig6bis}) 
 with respect to the optimal amplification $\omega_{\rm weak}=\omega^{(3)}$, i.e., in the case when
 the signal frequency is equal to the level splitting. Nevertheless, this amplification is still strong enough 
 ($\beta$ is about 20--30) allowing to use the proposed amplifier for a weak signal with arbitrary frequency.   

A similar situation occurs if the qubit coupling is not strong enough to reach the optimal regime (see Fig.~\ref{fig6bis}b).
Indeed, the amplification of a weak signal ($\epsilon/\Delta=0.5$) by the strong drive ($A/\Delta=12$) is weaker,
but still essential ($\beta$ about 10) for the coupling $g=0.1$. 

\section{Conclusions}

The spectrum of two coupled qubits driven by an ac signal with
the frequency in resonance with inter-level transitions has an unusual
structure, with a hierarchy of harmonic peaks with heights
non-monotonically dependent on the harmonic's number. This peak-height 
hierarchy is a fingerprint of any two-qubit system and can
be used to characterize both individual qubit parameters as well
as the interqubit coupling.

Exploiting the analogy between a parametric amplifier and a
system of two coupled qubits, we propose a method of amplification
of a weak signal via its mixing with a strong pump signal applied
to the two-qubit system. If both signals are relatively close to
the inter-level transitions in the four-level quantum system
(which can be achieved by tuning the qubit coupling), then the
amplification coefficient can be of the order of 100. 
When the weak signal frequency is different from the inter-level
splitting then the amplification is still strong enough, allowing the
proposed amplifier to work efficiently both in inter-level resonance and off 
inter-level resonance regimes. Weakening the qubit coupling also suppresses a
weaker signal enhancement, thus, requiring strongly-coupled qubits for this remarkable
parametric amplification. 

We also show that noise, which is of the order of a
weak signal, can strongly affect the time dependence of the output signal $Z_1(t)$,
but it modifies much weakly the spectrum $S_Z$. Therefore,
 the proposed amplifier can work efficiently in
noisy conditions.   
This large
amplification offers a different way of using multiqubit circuits
as parametric amplifiers.

\section{acknowledgments}

We acknowledge partial support from FRSF (Grant No. F28.21019) and
EPSRC (No. EP/D072518/1). FN acknowledges partial support from the
National Security Agency, Laboratory of Physical Science, Army
Research Office, National Science Foundation grant No.~0726909,
DARPA, AFOSR, JSPS-RFBR contract No.09-02-92114, MEXT Kakenhi on
Quantum Cybernetics,  and the JSPS-FIRST Program.

\section{appendix: Master equations for the density matrix components}

The master equation (\ref{master}) can be explicitly written as follows \cite{chem-phys}:
\begin{eqnarray}
\begin{array}{lll}
\dot{\Pi}_{0x}  & = & \Delta_2\Pi_{0y} - \Gamma_{\phi 2}\Pi_{0x} \\
\dot{\Pi}_{0y}  & = & -\Delta_2\Pi_{0x} + \epsilon_2(t)\Pi_{0z} - 2g\Pi_{xz} - \Gamma_{\phi 2}\Pi_{0y}\\
\dot{\Pi}_{0z}  & = & -\epsilon_2(t)\Pi_{0y} + 2g\Pi_{xy}-  \Gamma_{2}(\Pi_{0z}-Z_{T2})\\
& & \\
\dot{\Pi}_{x0}  & = & \Delta_1\Pi_{y0} - \Gamma_{\phi 1}\Pi_{x0} \\
\dot{\Pi}_{y0}  & = & -\Delta_1\Pi_{x0} + \epsilon_1(t)\Pi_{z0} - 2g\Pi_{zx} - \Gamma_{\phi 1}\Pi_{y0}\\
\dot{\Pi}_{z0}  & = & -\epsilon_1(t)\Pi_{y0} + 2g\Pi_{yx}-  \Gamma_{1}(\Pi_{z0}-Z_{T1})\\
& & \\
\dot{\Pi}_{xx}  & = & \Delta_2\Pi_{xy} + \Delta_1\Pi_{yx} - (\Gamma_{\phi 1} + \Gamma_{\phi 2})\Pi_{xx} \\
\dot{\Pi}_{xy}  & = & -2g\Pi_{0z} -\Delta_2\Pi_{xx} + \Delta_1\Pi_{yy} + \epsilon_2(t)\Pi_{xz} -    (\Gamma_{\phi 1} + \Gamma_{\phi 2})\Pi_{xy}\\
\dot{\Pi}_{yx}  & = & -2g\Pi_{z0} -\Delta_1\Pi_{xx} + \Delta_2\Pi_{yy} + \epsilon_1(t)\Pi_{xz} -    (\Gamma_{\phi 1} + \Gamma_{\phi 2})\Pi_{yx}\\
\dot{\Pi}_{xz}  & = & 2g\Pi_{0y} - \epsilon_2(t)\Pi_{xy} + \Delta_1\Pi_{yz} -  (\Gamma_{\phi 1}+\Gamma_{2})\Pi_{xz}\\
\dot{\Pi}_{zx}  & = & 2g\Pi_{y0} - \epsilon_1(t)\Pi_{yx} + \Delta_2\Pi_{zy} -  (\Gamma_{\phi 2}+\Gamma_{1})\Pi_{zx}\\
\dot{\Pi}_{yy}  & = & -\Delta_1\Pi_{xy} - \Delta_2\Pi_{yx} + \epsilon_2(t)\Pi_{yz} + \epsilon_1(t)\Pi_{zy} -  (\Gamma_{\phi 1} + \Gamma_{\phi 2})\Pi_{yy}\\
\dot{\Pi}_{yz}  & = & - \Delta_1\Pi_{xz} - \epsilon_2(t)\Pi_{yy} + \epsilon_1(t)\Pi_{zz} -  (\Gamma_{\phi 1}+\Gamma_{2})\Pi_{yz}\\
\dot{\Pi}_{zy}  & = & - \Delta_2\Pi_{zx} - \epsilon_1(t)\Pi_{yy} + \epsilon_2(t)\Pi_{zz} -  (\Gamma_{1}+\Gamma_{\phi 2})\Pi_{zy}\\
\dot{\Pi}_{zz}  & = & -\epsilon_1(t)\Pi_{yz} -\epsilon_2(t)\Pi_{zy}  -  (\Gamma_{1} + \Gamma_{2})(\Pi_{zz}-Z_{T1}Z_{T2})
\end{array}
    \label{poxy}
\end{eqnarray}
where the symbols are explained in Eqs.~(\ref{eq-ham}-\ref{eq:qomponents}).
In the absence of qubit-qubit coupling, $g=0$, the first three equations in (\ref{poxy}) describe the evolution of the Bloch vector components (${\Pi}_{0x}, {\Pi}_{0y}, {\Pi}_{0z}$)  of qubit 1, and the second three equations in (\ref{poxy}) describe those (${\Pi}_{x0}, {\Pi}_{y0}, {\Pi}_{z0}$) of  qubit 2.



\begin{thebibliography}{99}
\bibitem{Bellac} M. Le Bellac, {\em A Short Introduction to Quantum Information and Quantum Computation} (Cambridge University Press, 2006).
\bibitem{Nori-new} J.Q. You, F. Nori, Nature {\bf 474}, 589 (2011). 
\bibitem{Nori1}  J.Q. You and F. Nori, Physics Today {\bf 58}, No. 11, 42 (2005).
\bibitem{Nori2} I. Buluta, F. Nori, Science {\bf 326}, 108 (2009).
\bibitem{buluta} I. Buluta, S. Ashhab, F. Nori, Reports on Progress in Physics, in press (2011), arXiv:1002.1871.
\bibitem{smirnov} A.Yu. Smirnov, S. Savel'ev, L.G. Mourokh, F. Nori, Euro. Phys. Lett. {\bf 80}, 67008 (2007).
\bibitem{mi} A.M. Zagoskin, S. Savel'ev, F. Nori,
Phys. Rev. Lett. {\bf 98}, 120503 (2007).
\bibitem{metamat} A.L. Rakhmanov, A.M. Zagoskin, S. Savel'ev, and F. Nori, Phys. Rev. B {\bf 77}, 144507 (2008).
\bibitem{ost-zag} O. Astafiev, A. M. Zagoskin, A. A. Abdumalikov, Jr., Y. A. Pashkin, T. Yamamoto, K. Inomata, Y. Nakamura, and J. S. Tsai, Science 327, 840 (2010); A. A. Abdumalikov, O. Astafiev, A. M. Zagoskin, Yu. A. Pashkin, Y. Nakamura, 
J. S. Tsai, Phys. Rev. Lett. {\bf 104}, 193601 (2010).
\bibitem{Girvin2009} S. M. Girvin, M. H. Devoret and R. J. Schoelkopf, Phys. Scr. {\bf T137}, 014012 (2009).
\bibitem{Wendin2006} G. Wendin and V. S. Shumeiko, in: M. Rieth and W. Schommers (eds.),  {\em Handbook of Theoretical and Computational Nanotechnology}, v. 3 (American Scientific Publishers, 2006); also: arXiv:cond-mat/0508729.
\bibitem{Zagoskin2007}   A. Zagoskin and A. Blais, {\em Physics in Canada} {\bf 63}, 215 (2007); also: arXiv:0805.0164.
\bibitem{Mooij}  J.E. Mooij, T.P. Orlando, L. Levitov, L. Tian,
C.H. van der Wal, and S. Lloyd, Science {\bf 285}, 1036 (1999).
\bibitem{Orlando1999} T.P. Orlando, J.E. Mooij,   L. Tian,
C.H. van der Wal, L. Levitov, S. Lloyd, J.J. Mazo, Phys. Rev. B {\bf 60}, 15398 (1999).
\bibitem{Zagoskin2011} A.M. Zagoskin, {\em Quantum Engineering}, Cambridge University Press (2011).
\bibitem{param} V. Damgov, {\it Nonlinear and Parametric Phenomena: Theory and Applications in Radiophysical and Mechnical Systems} (World Scientific, Singapore, 2001).
\bibitem{molec} J.H. Posthumus, Reports on Progress in Physics {\bf 67}, 623 (2004).
\bibitem{cavities} C. Ciuti, P. Schwendimann, A. Quattropani, Semicond. Sci. Tech. {\bf 18}, S279 (2003).
\bibitem{squid} M. Muck, R. McDermott, Supercond. Sci. Tech. {\bf 23}, 093001 (2010).
\bibitem{Joseph} R. Vijay, M.H. Devoret, I. Siddiqi, Rev. Sci. Inst. {\bf 80}, 111101 (2009).
\bibitem{robert} J. R. Johansson, G. Johansson, C. M. Wilson, F. Nori, 
Phys. Rev. Lett. 103, 147003 (2009); Phys. Rev. A {\bf 82}, 052509 (2010). 
\bibitem{paul}  P. D. Nation, J. R. Johansson, M. P. Blencowe, F. Nori, arXiv:1103.0835. 
\bibitem{param-meta} A.D. Boardman, V.V. Grimalsky, Y.S. Kivshar, S.V. Koshevaya, M. Lapine, 
N.M. Litchinitser, V.N. Malnev, M. Noginov, Y.G. Rapoport, V.M. Shalaev, A.B. Kozyrev, D.W. van der Weide,
J. Phys. D - Applied Physics {\bf 41}, 173001 (2008).
\bibitem{quant-ampl} A.A. Clerk, M.H. Devoret, S.M. Girvin, F. Marquardt, R.J. Schoelkopf, Rev. Mod. Phys. {\bf 82}
1155 (2010). 
\bibitem{rakh} S. Savel'ev, A.L. Rakhmanov, and F. Nori, Phys. Rev. E {\bf 72}, 056136 (2005);  New J. Phys. {\bf 7}, 82 (2005).
\bibitem{nocoupl} M. Grajcar, A. Izmalkov, S.H.W. van der Ploeg, S. Linzen, E. Il'ichev, Th. Wagner, U. Hubner, H.-G. Meyer, Alec Maassen van den Brink, S. Uchaikin, and A.M. Zagoskin, Phys. Rev. B {\bf 72}, 020503(R) (2005).
\bibitem{coupl} S.H.W. van der Ploeg, A. Izmalkov, A. Maassen van den Brink, U. Huebner, M. Grajcar, E. Il'ichev, H.-G. Meyer, and A.M. Zagoskin,
 Phys. Rev. Lett. {\bf 98}, 057004 (2007).
 \bibitem{coupling-refs} M. Grajcar, Y.X. Liu, F. Nori, A.M. Zagoskin, Phys. Rev. B 74, 172505 (2006); S. Ashhab, S. Matsuo, N.
Hatakenaka, F. Nori, Phys. Rev. B 74, 184504 (2006); S. Ashhab,
A.O. Niskanen, K. Harrabi, Y. Nakamura, T. Picot, P.C. de Groot,
C.J.P.M. Harmans, J.E. Mooij, F. Nori, Phys. Rev. B 77, 014510
(2008).
\bibitem{you} J.Q. You, Y.X. Liu, C.P. Sun, F. Nori, Phys. Rev. B 75,
104516 (2007).
 \bibitem{Paauw} F.G. Paauw, A. Fedorov, C.J.P.M. Harmans, and J.E. Mooij, Phys. Rev. Lett. {\bf 102}, 090501 (2009).
\bibitem{i1} E. Il'ichev E, N. Oukhanski, A. Izmalkov, T. Wagner, M. Grajcar, H.-G. Meyer, A. Smirnov, A. Maassen van den Brink, M.H.S. Amin, and A.M. Zagoskin, Phys. Rev. Lett. {\bf 91}, 097906 (2003).
 \bibitem{mdbook} P. Moin, {\it Fundamentals of Engineering Numerical Analysis} (Cambridge University Press; 2010).
 \bibitem{ilich} E. Il'ichev, N. Oukhanski, T. Wagner, H.-G. Meyer, A. Yu.
Smirnov, M. Grajcar, A. Izmalkov, D. Born, W. Krech, and A.
Zagoskin, Low Temp. Phys. {\bf 30}, 620 (2004).
 \bibitem{chem-phys} A.N. Omelyanchouk, S. Savel'ev, A.M. Zagoskin, E. Il'ichev, F. Nori, Phys. Rev. B {\bf 80}, 
 212503 (2009); S. Savel'ev, A.M. Zagoskin, A.N. Omelyanchouk, F. Nori, Chem. Phys. 
{\bf 375} 180 (2010). 
\end{thebibliography}
\end{document}